\documentclass{osa-article}

\journal{osajournal}
\articletype{Research Article}
\usepackage{lineno}
\begin{document}

\title{Sub-minute Quantum Ghost Imaging in the infrared enabled by a "looking back" SPAD array}

\author{Valerio Flavio Gili,\authormark{1,*} Dupish Dupish,\authormark{1} Andres Vega,\authormark{1} Massimo Gandola,\authormark{2} Enrico Manuzzato,\authormark{2} Matteo Perenzoni,\authormark{2,°} Leonardo Gasparini,\authormark{2} Thomas Pertsch,\authormark{1,3} and Frank Setzpfandt\authormark{1,3}}

\address{\authormark{1} Friedrich-Schiller-Universität Jena, Institute of Applied Physics,
Abbe Center of Photonics, 07745 Jena, Germany.
\\
\authormark{2} Fondazione Bruno Kessler, 38123 Trento, Italy.
\\
\authormark{3} Fraunhofer Institute for Applied Optics and Precision Engineering, 07745 Jena,
Germany.
\\
\authormark{°} now at Sony Europe Technology Development Centre, 38123 Trento, Italy.

}

\email{\authormark{*}valerio.gili@uni-jena.de} %% email address is required



\begin{abstract}
 Quantum Ghost Imaging (QGI) is an intriguing imaging protocol that exploits photon-pair correlations stemming from spontaneous parametric down-conversion (SPDC). QGI retrieves images from two-path joint measurements, where single-path detection does not allow to reconstruct the target image. This technique, has been so far limited in terms of acquisition speed either by raster scanning, or by the slow electronics of intensified cameras. Here we report on a fast QGI implementation exploiting a SPAD array detector for the spatially resolving path, enabling the acquisition of a ghost image in under one minute. Moreover, the employment of non-degenerate SPDC allows to investigate samples at infrared wavelengths without the need for short-wave infrared (SWIR) cameras, while the spatial detection can be still performed in the visible region, where the more advanced silicon-based technology can be exploited. Our findings advance the state-of-the-art of QGI schemes towards practical applications. 
\end{abstract}

%%%%%%%%%%%%%%%%%%%%%%%%%%  body  %%%%%%%%%%%%%%%%%%%%%%%%%%
\section{Introduction}

Ghost Imaging (GI) is a widely investigated imaging protocol in which second-order light correlations are exploited to reconstruct the image of a target. In both its classical and quantum versions, light is sent through two paths, one of which contains the target and a single detector with large numerical aperture (NA), i.e. the bucket detector, collecting all light from the target, while the second path contains a spatially resolving detector. Only a correlation reconstruction from the two detector measurements allows to retrieve the target image, while the two individual measurements do not contain enough information \cite{padgett-review}. While at first GI was believed to be achievable only by quantum entangled photon-pairs generated through spontaneous parametric down-conversion \cite{klyshko, shih}, it was later discovered that classical correlations of light enable this technique as well, and a number of experimental demonstrations followed \cite{usIEEE}, exploiting angle-correlated pulses \cite{ClGI}, speckle correlations of pseudo-thermal sources \cite{shihCl}, or true thermal sources \cite{thermalGI}. Despite the large number of possible implementations, quantum ghost imaging (QGI) promises better performances than its classical counterpart in terms of the signal-to-noise ratio (SNR) \cite{shap-boyd, meda}. Moreover, QGI performed with non-degenerate SPDC photon-pairs allows for two-color schemes, such that targets can be investigated at wavelengths where single-photon cameras are expensive or not yet available, such as the infrared \cite{twocolor, twocolor2}, or the terahertz \cite{teragi}, while the spatial detection can be still performed in the visible, where the well developed silicon-based technology can be exploited. Since the first demonstration of QGI, which relied on raster scanning in the spatial detection path, efforts have been made to advance the implementation of QGI towards applications, exploiting technological advances \cite{marta}. The reason lies in an inherent limit in detection efficiency in a raster scanned system to $1/N$, with $N$ the number of scanned pixels \cite{ICCD1}. An important technical leap was enabled by the employment of intensified charge-coupled device cameras (ICCD) for the spatial detection \cite{ICCD1, twocolor, isnp}, in which the detection efficiency becomes proportional to the number of pixels $N$. However, time correlations between bucket detector signals and individual pixels are not possible in such devices, and hence correlations with individual frames are necessary to reconstruct a ghost image. To do so, gating of the ICCD intensifier with bucket detector signal arrivals is necessary, together with a photon sparsity requirement, ensuring that at most one photon is detected per gate duration. Such requirements on time-correlation reconstruction result in typical frame rates no larger than $\sim$kHz, and on the other hand require to preserve the quantum state of idler photons propagating towards the ICCD camera with the use of $\sim$20-30 m long image-preserving delay lines, which account for the electronic delay between bucket detector electric signal arrival and ICCD intensifier gating mechanism. A way to rule out the need for such bulky delay lines is to exploit entanglement between a photon and a quantum memory in magnetic-trapped, laser-cooled atom systems \cite{varsavia}, which however introduces further technical difficulties, and does not solve the intrinsic frame rate limitations of ICCDs. More recently, Single Photon Avalanche Diode (SPAD) arrays have reached a high technological maturity \cite{spadreview}, featuring increased observation rates up to 1 MHz, $\sim$200 ps resolution\cite{FBK}, close-to 100\% fill factor, and megapixel resolution \cite{3dspad}, and various detection operations such as timstamping, counting, and gating. Such achievements have brought about a tremendous state-of-the-art advancement in a number of bio and quantum imaging applications, ranging from near-infrared (NIR) fluorescence lifetime imaging \cite{bioSPAD}, pixel super resolution \cite{supres}, quantum optical centroid measurements \cite{centr}, EPR inequality characterization \cite{eprspad}, correlation plenoptic imaging \cite{plenspad}, Hong-Ou-Mandel interference microscopy \cite{spadhom}, ranging \cite{spadlidar}. In parallel, single-photon cameras based on the so-called "Timepix" technology have been developed as well, enabling results such as Hong-Ou-Mandel interference \cite{timepix}, 3D imaging \cite{timepixcpi}, and hyperspectral quantum imaging \cite{timepixhyp}. The advantage of using a SPAD array lies in the possibility to have in-pixels TDCs, bringing about the ability to reconstruct pixel-wise time correlations with bucket detector signals without the need for image-preserving delay lines. An attempt to employ such technology to advance the state-of-the-art of QGI has been recently made, even though it is so far limited in speed by the employment of a SPAD line, rather than a 2D array, which requires detector raster-scanning to reconstruct 2D images \cite{SPADGI}, thus fundamentally limiting the acquisition speed and detection efficiency along one axis. In our work, we demonstrate fast QGI using a 2D SPAD array, enabling to construct ghost images in under one minute acquisition time, with a frame rate $\sim$35 kHz, about one order of magnitude faster than previous demonstrations relying on ICCD technology. The employment of in-pixel TDCs enables the reconstruction of pixel-wise time correlations with bucket detector signals and avoids the necessity of bulky delay lines. Time correlations are reconstructed by employing the so-called "looking back" principle to synchronize SPAD array and bucket detector, in which idler photon detection events act as a STOP signal, with in-pixels TDCs providing enough electronic delay to build-up pixel-wise correlation histograms within a range of 100 ns, hence removing the need of image-preserving delay lines. Importantly, the frame rate is mainly limited by the idler photon flux, i.e. the detection efficiency and dead time of the NIR bucket detector, since in principle our zero-suppression circuit speeding up the read-out process, could enable observation rates up to 1 MHz. In line with previous demonstrations \cite{twocolor, twocolor2}, our implementation also allows for two-color imaging, and we demonstrate this by performing QGI of a Si-Au sample at 1.4 $\mu$m wavelength, by using our Si SPAD array device for the spatial detection, which would not be able to resolve such sample in direct imaging configurations without a SWIR camera. Our results advance the state-of-the-art of QGI towards applications and further confirm the recent advancement of quantum technologies enabled by the development of SPAD array and Timepix cameras.

\section{Setup and detector synchronization}

\begin{figure}[h!]
\centering\includegraphics[width=\textwidth,keepaspectratio]{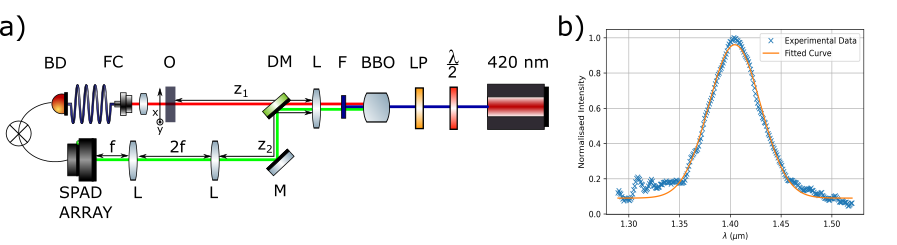}
\caption{a) Schematics of the setup for QGI: a 420 nm pump illuminates a Beta-Barium Borate (BBO) crystal after polarization tuning with a $\lambda$/2 waveplate and a linear polarizer (LP). A lens (L) collimates photon pairs, after a longpass filter (F) cuts out any residual pump. A dichroic mirror (DM) separates photons according to their wavelength; the VIS photon goes through a telescope system of 2 lenses and gets detected by a SPAD array imager. The NIR photon encounters an object (O) placed on a x-y translation stage, and gets detected with an InGaAs bucket detector (BD) after passing through a fiber coupling stage (FC). Finally, signals from NIR photon detection events are sent to the same FPGA controlling the SPAD array to reconstruct pixel-by-pixel quantum correlations. b) Spectrum of the NIR idler photon centered at 1.4 $\mu$m, corresponding to a signal photon wavelength at 600 nm, acquired with an InGaAs spectrometer.}
\end{figure}

The setup we used to demonstrate fast QGI is depicted in Fig. 1a. A Continuous Wave (CW) pump centered at 420 nm with 480 mW power impinges on a Beta-Barium Borate (BBO) $\chi^{(2)}$ nonlinear crystal cut for Type-I SPDC process, to produce non-degenerate photon pairs, such that one photon is in the visible (VIS) range and one photon is in the near-infrared (NIR). The NIR idler photon spectrum is measured with an InGaAs spectrometer (Fig. 1b), showing a peak at 1.4 $\mu$m wavelength, such that the signal photon wavelength can be retrieved from the energy conservation condition in SPDC: $\lambda_s^{-1}=\lambda_p^{-1}-\lambda_i^{-1}$, from which a $\lambda_s=600$ nm can be inferred. A $\lambda /2$ waveplate and a linear polarizer (LP) are used to control the pump polarization state and ensure the desired phase-matching condition is fulfilled. After pump filtering, a lens (L) is used to collimate signal and idler cones, such as to have the BBO crystal plane at infinity.
%an imaging condition can then be found as long as the distances z$_1$, z$_2$ from the collimating lens to the object and spatial detector are equal, regardless of their magnitude. 
The photon pairs are then separated in wavelength with a dichroic mirror (DM): the signal photons at 600 nm go through a telescope system, to magnify the single-photon spot, before being detected by our SPAD array, while idler photons at 1.4 $\mu$m go through an object (O) mounted on a x-y translation stage, get fiber coupled and finally detected by a low-dark-count air-cooled InGaAs detector (ID 230, ID Quantique), effectively acting as a "bucket detector" with no spatial resolution. An imaging condition can be found using the unfolded version of the QGI scheme, the so-called "Klyshko advanced-waves picture" \cite{klyshko2}, as long as the distance $z_1$ from the collimating lens after the BBO cystal and the object in the NIR arm matches distance $z_2$ from the collimating lens to the first lens of the telescope system in the VIS arm. Signals coming from NIR photon detection events are sent directly to an FPGA controlling the SPAD array to perform on-chip correlation measurements. 

\begin{figure}[h!]
\centering\includegraphics[width=\textwidth,keepaspectratio]{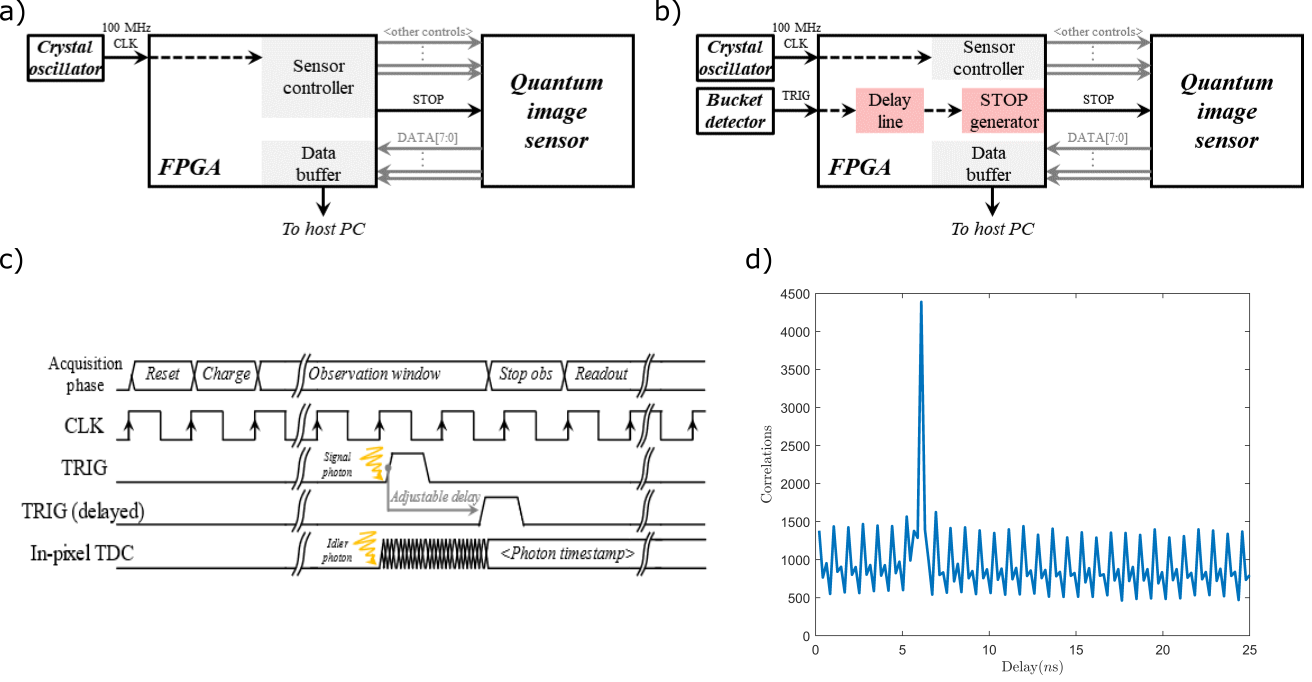}
\caption{a) SPAD array acquisition mechanism in a direct imaging configuration: a crystal oscillator attached to a FPGA sets the time basis for all in-pixel TDCs. b) "Looking back" acquisition mechanism for our QGI experiment: after any pixel of the SPAD array detects a photon, an observation window is opened, such that the reception of a trigger from the bucket detector acts as a STOP signal. c) Time diagram of the acquisition process. d) Typical correlation histogram acquired for a single pixel.}
\end{figure}

The single-photon time-resolved SPAD sensor consists of 32×32 pixels with 44.54 $\mu$m pitch, specifically designed for quantum imaging applications and manufactured in 150 nm CMOS standard technology \cite{FBK}. Each pixel, featuring 19.5\% fill-factor, contains a SPAD, having a peak Photon Detection Probability (PDP) of $\sim$0.3 at 3 V excess bias, with its quenching and front-end circuits, and a 8-bit time-to-digital converter (TDC) with about 200 ps resolution and the readout electronics. An in-pixel 1-bit memory can disable a noisy pixel to reduce the accidental correlations.
The in-pixel TDC is based on a gated ring oscillator clocking a counter; the ring starts upon detection of a photon and stops when the globally distributed STOP signal is activated. In a typical configuration \cite{eprspad}, the acquisition system is synchronous with an FPGA-based primary unit, which defines the time-base synchronously with a 100 MHz clock provided by a low-jitter crystal oscillator attached to the FPGA itself (see Fig. 2a). This configuration has been specifically modified to comply with the need of finding a common time-reference with the signals coming from the NIR bucket detector, which are asynchronous with the system clock, to retrieve photon-pair correlations. In our ghost imaging detector configuration, sketched in Fig. 2b, the STOP signal is generated synchronously with the TRIG signal generated by the bucket detector. An electronic programmable delay line implemented in the FPGA compensates for delays due to optical path differences up to 100 ns; viceversa, the in-pixel TDC working in reverse START-STOP mode provides enough range (50 ns) to isolate the photon coincidences with no need for optical delay lines in the signal path \cite{twocolor, isnp}. 
Dedicated zero-suppression circuits implementing row-skipping and frame-skipping working modes on chip reduce the frame occupancy implemented on-chip with the aim to increase the duty cycle. While no detection event is observed, the chip works in frame-skipping mode, discarding frames without read-out operation; this mechanism allows a maximum observation rate of 1 MHz. When a trigger photon gets detected, the row-skipping modality comes into play: read-out is performed one row at a time skipping empty rows. In this case, assuming one event is detected on a single row, the maximum allowed observation rate is 0.31 MHz. Figure 2c shows a simplified timing diagram of the acquisition process: after resetting the pixel, all the SPADs in the array are synchronously charged, defining the beginning of the observation window. The window is kept indefinitely opened until a trigger from the bucket detector is received, acting as the STOP. Then, the readout procedure is performed, transferring all pixel values to the FPGA before being transmitted to a host PC via a USB 3.0. A LabView interface on the host PC continuously registers the stream of correlations as a function of time delay, building up over time a correlation histogram for every pixel, effectively "looking-back" in the stream of visible photon detection events, to find correlations with NIR photons. A typical correlation histogram is shown in Fig. 2d, showing a zero-delay correlation peak for an electronic delay of 5.9 ns. Once data collection is over, the individual pixel histograms can be saved in ASCII format and post processed to remove the background and build the ghost image. Importantly, the background of the correlation histogram shows a deterministic pattern due to transistor mismatch in the ring oscillator in the TDC, such that bins can be in practice longer of shorter than 200 ps. The pattern period is four bins, and coincides with the four phases of the ring oscillator. This pattern can be characterized and calibrated in post-processing or removed with a moving average; the latter however causes a loss in resolution.   

\section{Ghost Imaging experiment}

To demonstrate ghost imaging with our setup, we characterized a USAF negative resolution target (Thorlabs R3L3S1N, Fig. 3), in the sample area labelled by group number "2", element numbers "4", "5", and "6", corresponding to a slit width of 1.41mm, 1.26 mm, and 1.12 mm, respectively. We investigated four different areas of the mentioned area, labelled in Fig. 3 by colored circles on the USAF sample scheme, corresponding to ghost images with matching frame colors. The four ghost images shown in Fig. 3 were constructed by accumulating photon counts for $\sim 20$ min, followed by exporting correlation histograms for every detector pixel, as in Fig. 2d. After background removal, and correlation peak integration, the total number of coincidences per pixel can be estimated, and a 2D ghost image could be reconstructed. 

\begin{figure}[h!]
\centering\includegraphics[width=\textwidth,keepaspectratio]{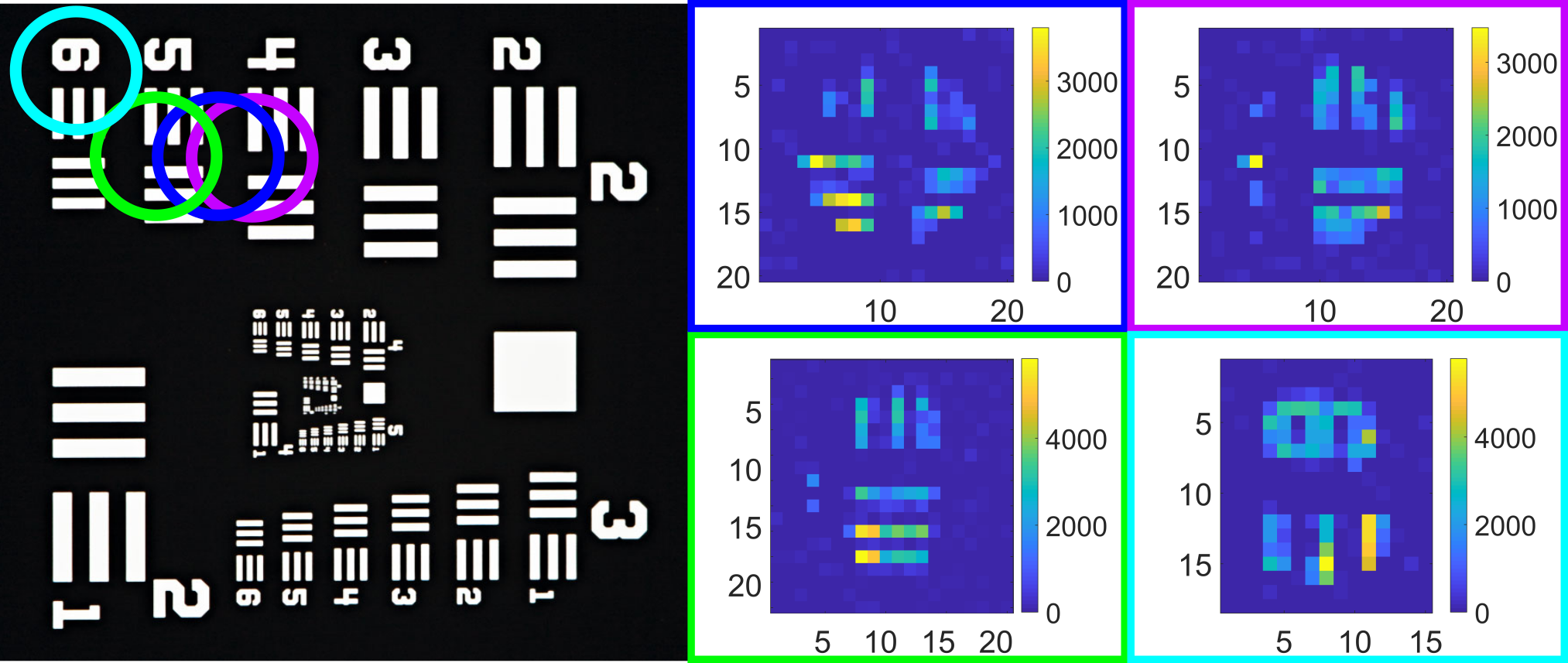}
\caption{QGI of a negative USAF resolution target. Left: schematics of the object, with four specific zones highlighted with colored circles. Every circled zone corresponds to one of the four ghost images shown in the right side, with frame colors matching circle colors. In every ghost image x and y axis represent the pixel number, and the colormap indicates the number of correlations. }
\end{figure}

After a first characterization of our object, we proceeded to investigate the speed of our QGI configuration, we selected one area of the USAF target (Fig. 3, green frame/circle) and varied photon accumulation times in the range 10 s - 20 min, while keeping our pump power fixed at 480 mW; this in turn corresponded to a NIR photon flux of 73 kHz at a detection efficiency of 25\%, and dead time of 8 $\mu$s; such NIR photon flux corresponds to typical observation rates of $\sim$ 35 kHz. This value could be increased by reducing the bucket detector dead time, whose minimum reaches 2 $\mu$s. This would however negatively impact the image SNR due to the heavily increased NIR dark count rate. Figure 4 reports the QGI results obtained for a selection of accumulation times, below one minute (top row), and between 1 min and 20 min (bottom row). Overall, the images obtained for $\le$1 min accumulation show that object features can be clearly appreciated after 20 s - 30s (Fig. 4b,c), while the image at 60 s (Fig. 4e) contains all sample features, albeit more noisy compared to the image obtained after 20 min (Fig. 4j) experiment duration. Importantly, all accumulation times here mentioned refer to the total experiment duration, while the effective exposure time is actually smaller because of the influence of the read-out time. Potentially, this could be optimised in the future to make the actual exposure time as close as possible to the actual experiment duration. 

\begin{figure}[h!]
\centering\includegraphics[width=\textwidth,keepaspectratio]{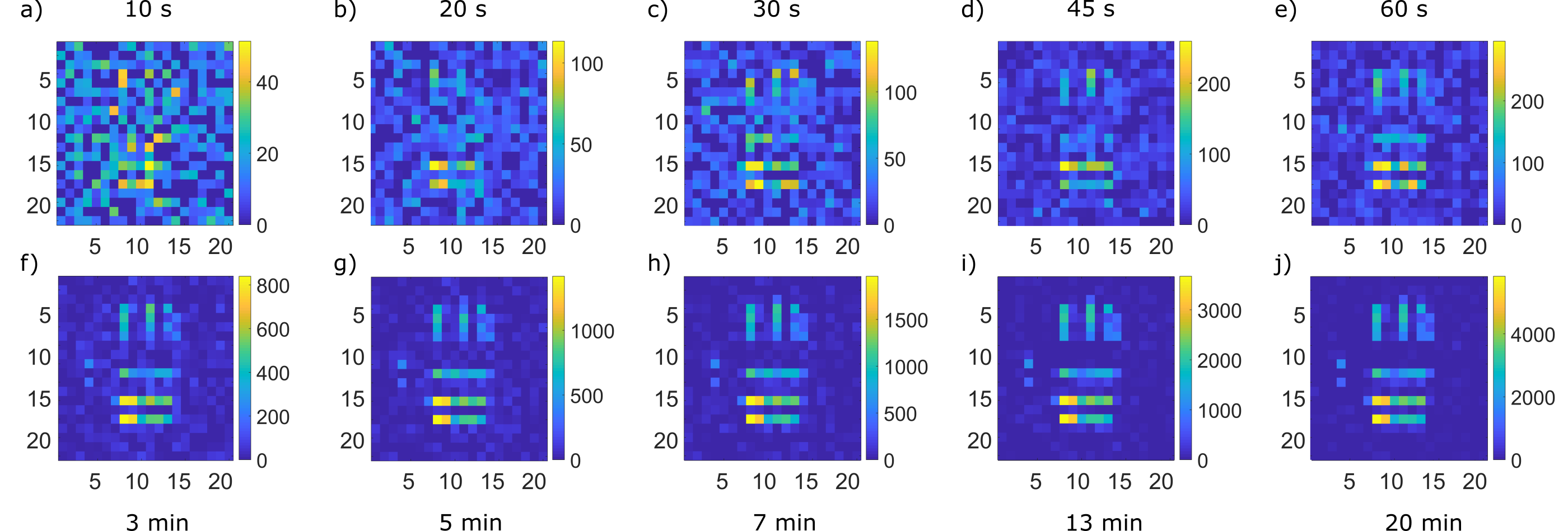}
\caption{a)-j) QGI of the same USAF target area of Fig. 3, green circle, obtained for different integration times. Top row, from left to right: a) 10 s, b) 20 s, c) 30 s, d) 45 s, e) 60 s. Bottom row, from left to right: f) 3 min, g) 5 min, h) 7 min, i) 13 min, j) 20 min. For each image, x and y axes indicate the pixel number, while the colormap indicates the number of correlations. }
\end{figure}

Finally, to fully demonstrate the two-color capabilities of our scheme, we performed QGI of a Si-Au target. The choice of such materials is motivated by the fact that at visible wavelengths no image of such structures would be possible, since Si would absorb and gold would reflect, while in the NIR Si is transparent. Hence, it is possible to inspect such object in the NIR, while still performing the spatially resolved detection with a Si imager device \cite{twocolor}. To this end, a sample consisting of 2 mm wide Au slits resting upon a double-polished Si wafer was fabricated using a standard photolithography process, followed by 50 nm Au deposition and lift-off. This choice of the Au layer thickness is motivated by the fact that 50 nm of Au exhibit a reflectance >95 \% at 1.4 $\mu$m, estimated with Fresnel equations. The experimental results are reported in Fig. 5: a camera picture of the fabricated sample is shown to the left, while to the right two exemplary ghost images are reported, acquired in 2 min and 5 min, corresponding to the two circles in the same panel with matching color compared to the image frames. In contrast with the USAF target, where bright (transmissive) areas correspond to the glass substrate and dark (reflective) areas correspond to chrome, here dark areas correspond to Au structures, whereas bright areas indicate the transmissive Si wafer.

\begin{figure}[h!]
\centering\includegraphics[width=\textwidth,keepaspectratio]{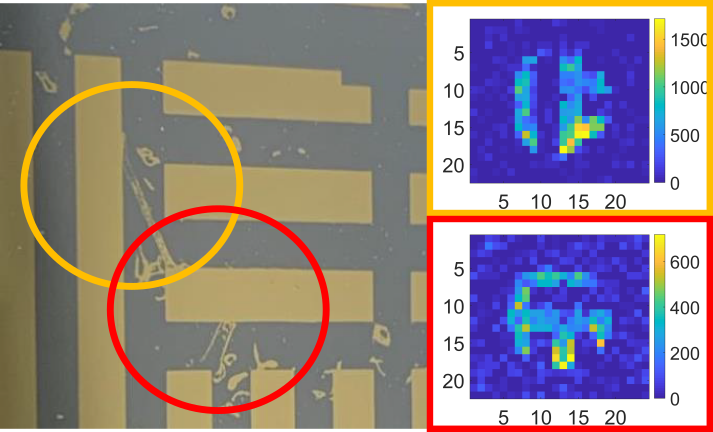}
\caption{QGI of a Si-Au target consisting of 2 mm wide Au slits on a Si substrate. Left: camera picture of the sample, with circles corresponding to the areas imaged in the two ghost images on the right side. Circle colors match frame colors of same areas. Accumulation times for these images amounted to $\sim$5 min (yellow frame), and  $\sim$2 min (red frame).}
\end{figure}

\section{Conclusion}
With this work, we proposed a compact setup to realise QGI in the infrared. The core of our advancements lies with the fast electronic processing of our SPAD array imager, with in-pixel TDCs clocked by an external FPGA, which also synchronises detection events coming from our NIR bucket detector. Thanks to this device, we were able to perform fast QGI without the need for bulky delay lines \cite{isnp}, or complex quantum memory systems \cite{varsavia}, with accumulation times lower than 1 min. We further demonstrated the two-color imaging capabilities of QGI of a Si-Au object \cite{twocolor}, which would not be possible in direct imaging configurations with a Si imager. Our results push forward the state-of-the-art of QGI implementations, thanks to the rapidly growing of SPAD array technology, and point towards further developments of the two-color feature towards the MIR, where the fingerprint region is of particular interest for chemical industry, and low light illumination, combined with the robustness of correlations against turbulence \cite{turbolence}, and turbid \cite{turbid} or scattering media \cite{scat}, can be exploited for practical applications.

\begin{backmatter}
\bmsection{Funding}
This work was supported by the Thuringian Ministry for Economy, Science, and Digital Society and the European Social Funds (2021 FGI 0043); European Union’s Horizon 2020 research and innovation programme (Grant Agreement No. 899580); the German Federal Ministry of Education and Research (FKZ 13N14877, 13N15956); and the Cluster of Excellence ‘‘Balance of the Microverse’’ (EXC 2051 – project 390713860). 

\bmsection{Acknowledgments} V. F. Gili thanks I Staude, M. Steinert, M. Rikers, and J. Gour for help in the fabrication of the Si-Au sample, and M. Gräfe for fruitful discussions.

\bmsection{Disclosures} The authors declare no conflicts of interest.

\bmsection{Data Availability Statement} Data underlying the results presented in this paper are not publicly available at this time but may be obtained from the authors upon reasonable request.

%\bmsection{Supplemental document}
%See Supplement 1 for supporting content. 

\end{backmatter}

%%%%%%%%%%%%%%%%%%%%%%% References %%%%%%%%%%%%%%%%%%%%%%%%%

%%%%%%%%%% If using BibTeX:
\bibliography{sample}

%%%%%%%%%% If preparing manually:
% \newcommand{\enquote}[1]{``#1''}

% \bibitem{Zhang:14}
% Y.~Zhang, S.~Qiao, L.~Sun, Q.~W. Shi, W.~Huang, L.~Li, and Z.~Yang,
%   \enquote{Photoinduced active terahertz metamaterials with nanostructured
%   vanadium dioxide film deposited by sol-gel method,}
%   {\protect\JournalTitle{Optics Express}} \textbf{22}, 11070--11078 (2014).

% {Optical Society}, \enquote{{OSA Publishing},}
%   \url{http://www.osapublishing.org}.

% \bibitem{FORSTER2007}
% P.~Forster, V.~Ramaswamy, P.~Artaxo, T.~Bernsten, R.~Betts, D.~Fahey,
%   J.~Haywood, J.~Lean, D.~Lowe, G.~Myhre, J.~Nganga, R.~Prinn, G.~Raga,
%   M.~Schulz, and R.~V. Dorland, \enquote{Changes in atmospheric consituents and
%   in radiative forcing,} in \enquote{Climate Change 2007: The Physical Science
%   Basis. Contribution of Working Group 1 to the Fourth assesment report of
%   Intergovernmental Panel on Climate Change,}  S.~Solomon, D.~Qin, M.~Manning,
%   Z.~Chen, M.~Marquis, K.~B. Averyt, M.~Tignor, and H.~L. Miler, eds.
%   (Cambridge University Press, 2007).

% \end{thebibliography}

\end{document}